\documentclass[twocolumn,english,aps,prb,floats]{revtex4}
\usepackage[T1]{fontenc}
\usepackage[latin9]{inputenc}
\usepackage{babel}
\usepackage{bm}
\usepackage{amsmath}
\usepackage{amssymb}
\usepackage{graphicx}
\usepackage{wasysym}
\usepackage[toc,page]{appendix}
\usepackage{esint}
\usepackage{float}
\usepackage[unicode=true,
 bookmarks=false,
 breaklinks=false,pdfborder={0 0 1},backref=section,colorlinks=false]
 {hyperref}
\usepackage{breakurl}

\makeatletter
\@ifundefined{textcolor}{}
{%
 \definecolor{BLACK}{gray}{0}
 \definecolor{WHITE}{gray}{1}
 \definecolor{RED}{rgb}{1,0,0}
 \definecolor{GREEN}{rgb}{0,1,0}
 \definecolor{BLUE}{rgb}{0,0,1}
 \definecolor{CYAN}{cmyk}{1,0,0,0}
 \definecolor{MAGENTA}{cmyk}{0,1,0,0}
 \definecolor{YELLOW}{cmyk}{0,0,1,0}
}

\@ifundefined{textcolor}{}{%
 \definecolor{BLACK}{gray}{0}
 \definecolor{WHITE}{gray}{1}
 \definecolor{RED}{rgb}{1,0,0}
 \definecolor{GREEN}{rgb}{0,1,0}
 \definecolor{BLUE}{rgb}{0,0,1}
 \definecolor{CYAN}{cmyk}{1,0,0,0}
 \definecolor{MAGENTA}{cmyk}{0,1,0,0}
 \definecolor{YELLOW}{cmyk}{0,0,1,0}
}

\usepackage{babel}
\usepackage{ifpdf}\usepackage{bm}

\makeatother

\begin{document}

\title{Elliptical skyrmions: theory and nucleation by a magnetic tip in an antiskyrmion-hosting material}

\author{Daniel Capic}
\email{dcapic11@gmail.com}
\affiliation{The College of New Jersey, 2000 Pennington Road, Ewing, New Jersey, 08618 USA }

\date{\today}
\begin{abstract}
Motivated by experimental findings in [ \onlinecite{JenaElliptical} ] and [ \onlinecite{PengTransformation} ], we find an elliptical skyrmion can be nucleated in a material with a Dzyaloshinskii-Moriya interaction (DMI) that supports antiskyrmions. The DMI favors the elongation of the skyrmion along a given direction so that it is characterized by two size parameters. We derive analytical properties of the elliptical skyrmion using the Belavin-Polyakov (BP) pure exchange model. We show that the DMI which typically favors isotropic skyrmions or antiskyrmions, respectively, can also support elliptical antiskyrmions or skyrmions, respectively. Using the real material parameters extracted from \cite{JenaElliptical} and including all relevant interactions, numerical computations indicate that the elliptical skyrmion can be nucleated from the stripe or labyrinth domain state by a magnetic force microscope tip in a thin film with a DMI favoring antiskyrmions that consists of at least a few atomic layers. 

\end{abstract}

\maketitle

\section{Introduction}
Skyrmions are topologically non-trivial solutions that occur in non-linear field theories \cite{Skyrme1958}. In the context of magnetic systems, skyrmions appear as a defect in the ferromagnetic background. Skyrmions are an emerging area of research due to their potential for computing applications, see for instance the review by Fert \textit{et al.} \cite{FertReview} or Zhang \textit{et al.} \cite{Zhang2020}.

The stability of skyrmions in various crystal symmetries has been studied in some detail, see for instance \cite{BogdanovVortices}, \cite{BogdanovPioneer}. The skyrmion can be stabilized by the anisotropic exchange interaction, also referred to as the Dzyaloshinskii-Moriya interaction (DMI) \cite{DzialoshinskiiWeakFerromagnetism} \cite{MoriyaAnisotropic}. In a general form, the interaction can be written in terms of the spin vector $\mathbf{s}$ as:
\begin{equation}
E_{DMI} = \frac{1}{a}\int d^2\rho \; \mathbf{D}_{ij} \cdot (\mathbf{s}_{i} \times \mathbf{s}_{j})
\end{equation}
where $\mathbf{D}_{ij}$ is the Dzyaloshinskii vector whose direction is dictated by the crystal symmetry. Materials with inversion symmetry have $\mathbf{D}_{ij}=0$.  The crystal symmetry can be described using \cite{DLifshitz}:
\begin{equation}
{\cal L}^{(k)}_{ij}= s_{i} \frac{\partial s_{j}}{\partial r_{k}} -  s_{j} \frac{\partial s_{i}}{\partial r_{k}}
\end{equation}
i.e. the Lifshitz invariants \cite{Lifshitz}. In a thin film with an isotropic DMI, where the DMI vector is the same magnitude along any given direction, see for instance \cite{HuangAnisotropic}, materials that have the "N{\'e}el-type" or "Bloch-type" DMI have the corresponding DMI energy density \cite{BogdanovVortices} $D({\cal L}^{(y)}_{xz}+{\cal L}^{(x)}_{zy})$ and  $D({\cal L}^{(y)}_{xz}-{\cal L}^{(x)}_{yz})$. Such systems will support "Neel-type" and "Bloch-type"  skyrmions, respectively. 

Typically, skyrmions stabilized by the DMI have an isotropic circular shape, see for instance \cite{RosslerSpontaneous} \cite{MuhlbauerDiscovery} \cite{MunzerLattice} \cite{HanLattice} \cite{YuFeGe} \cite{HeinzeSpontaneous}. When the DMI is isotropic, Dujin \textit{et al.} \cite{DuijnSTT} have found that Bloch skyrmions elongated by the spin-transfer torque returned to the isotropic circular state. Lin \textit{et al.} \cite{LinInternal} showed that skyrmions can be deformed depending upon the skyrmion's internal modes. The material parameters used by Chui \textit{et al.} \cite{ChuiGeometrical} yielded nearly circular skyrmions when confined to a nanowire, while Kim \textit{et al.} \cite{KimCircular} and Yang \textit{et al.} \cite{YangModes} found that a skyrmion confined to a circular thin magnetic dot and magnetic nanotube, respectively, could deform into an elliptical shape, suggesting skyrmion confinement can also lead to distortions.  Mokkath \cite{Mokkarth} predicted that elliptical skyrmions can emerge in a bilayer at lower applied fields. Lin and Saxena \cite{LinOblique} showed that a skyrmion in a thin film can be deformed by applying a tilted magetic field.  Derras-Chouk and Chudnovsky \cite{ChoukDefects} found that defects can also lead to the deformation of skyrmions. If the system contains an elliptical skyrmion, M{\"u}ller \textit{et al}. \cite{MullerSaddle} have shown analytically that it can divide into a pair of skyrmions. 

The skyrmion can also be deformed by altering the material parameters of the system. Koibuchi \textit{et al.} \cite{KoibuchiShape} calculated the skyrmion deformation due to mechanical stresses that modulate the DMI, while Hu \textit{et al.} \cite{HuStrain} have seen that elliptical skyrmions can be induced when there is an anisotropic strain. Mehmood \textit{et al.} \cite{MehmoodPMA} showed that the skyrmion shape can be manipulated by changing the perpendicular magnetic anisotropy. Garanin \textit{et al.} \cite{GaraninThermal} observed a few elliptical skyrmions in a numerical experiment produced by freezing a system of labyrinth domains down to zero temeperature. 

If one considers a model with the anisotropic DMI, e.g. $D_{x} \neq D_{y}$, Udalov \textit{et al.} \cite{UdalovDMI} have shown numerically and analytically that skyrmions and antiskyrmions deform into elliptical shapes. In a related area, Osorio \textit{et al.} \cite{OsorioGroup} and Huang \textit{et al.} \cite{HuangAnti} have shown that an antiferromagnetic skyrmion can be stabilized by the anisotropic DMI which distorts the skyrmion into the same elliptical shape as the ferromagnetic skyrmion already discussed. Perhaps relevant for future works, Kuchkin and Kiselev \cite{KuchkinChiral} have shown that the equilibrium antiskyrmion configuration in a material with the DMI that supports skyrmions should be elliptical.

Experimentally, in two works, Camosi \textit{et al.} \cite{CamosiDMI} \cite{CamosiElliptical} demonstrated that Au/Co/W(110) with the anisotropic DMI due to the $C_{2v}$ crystal symmetry could support isolated elliptical skyrmions. Wilson \textit{et al.} \cite{WilsonLamallae} observed the distorted elliptical lattice state in the material Cu$_2$OSeO$_3$ when the applied magnetic field was tilted with respect to the direction normal to the film.  Cui \textit{et al.} \cite{CuiMultilayer} experimentally observed N{\'e}el-type elliptical skyrmions when they obliquely deposited one layer into a multilayer system.  

It seems advantageous to study elliptical skyrmions in the context of high-speed skyrmion motion. Kuchkin \textit{et al.}\cite{KuchkinSkyrmionDynamics} have found that elliptical skyrmions can be driven at higher speeds than isotropic skyrmions, so that elliptical skyrmions seem more attractive for high-speed computing applications. Psaroudaki and Panagopoulos \cite{PsaroudakiQubits} have shown that the elliptical skyrmion can be used to realize skyrmion qubits. Xia \textit{et al.} \cite{XiaEllipticalDynamics} have found that the skyrmion Hall angle, which describes the deflection of the skyrmion, can be adjusted based on the elliptical deformation of the skyrmion.  Masell \textit{et al.} \cite{Masell STT} and Fernandes \textit{et al.} \cite{FernandesImpurity} have shown that skyrmions driven by a large spin-polarized current deform into elliptical shapes. Litzius \textit{et al.} \cite{LitziusDynamics} have found that the maximum skyrmion velocity when driven by an electric current depends on the deformation and the temperature.  Fook \textit{et al.} \cite{FookTransport} found that skyrmions can be compressed into distorted elliptical shapes on the application of an electric or magnetic field and that these compressed skyrmions could be depinned from impurities by a lower current density and driven in a single direction. If a temperature gradient is applied, Wang \textit{et al.} \cite{WangRectilinear} have found that the deformation of the skyrmion drives its motion, while Kerber \textit{et al.} \cite{KerberAnisotropic} found that the diffusion of skyrmions deformed by an in-plane field depended on their elliptical deformation. 

Antiskyrmions are related objects with a different topological index. As shown by Bogdanov \textit{et al.} \cite{BogdanovPioneer}, antiskyrmions can be stabilized by structures with the $D_{2d}$ crystal symmetry. Antiskyrmions have been observed experimentally in Heusler compounds that have this symmetry \cite{JenaElliptical} \cite{PengTransformation} \cite{NayakTetragonal}\cite{SahaAnti} \cite{KumarTHE} \cite{MaHeussler} \cite{YasinBlochLines} \cite{JenaFerrimagnet}. Karube \textit{et al.} \cite{KarubeSawtooth} have also observed antiskyrmions in a chiral magnet with $S_{4}$ symmetry. 
Pierobon \textit{et al.} \cite{PierobonPhase} have shown numerically that the application of a magnetic field to a skyrmion lattice can produce an "inverted" lattice phase by producing antiskyrmions. Directly relevant to this work, Jena \textit{et al.} \cite{JenaElliptical} and Peng \cite{PengTransformation} have observed both antiskyrmions and elliptically distorted skyrmions in the material Mn$_{1.4}$Pt$_{0.9}$Pd$_{0.1}$Sn which also has the $D_{2d}$ crystal symmetry. In another work, Jena \textit{et al.} \cite{JenaChiral} observed antiskyrmions and elliptical skyrmions in a nanostripe of the same material. 

It has been shown by S. Zhang \textit{et al.} \cite{SenfuZhangMFMDomainWalls} that a magnetic force microscope (MFM) can be used to write an isotropic skyrmion from the domain state. Later, Garanin \textit{et al.} \cite{GaraninMFM} showed that it is theoretically possible for a MFM tip to write the isotropic skyrmion from the uniform ferromagnetic state in non-chiral thin films. Recently, Sharma \textit{et al.} \cite{SharmaMFM} have found that a collection of nano-objects can be written from a thin film with the $D_{2d}$ symmetry in Mn$_{2}$RhSn in the absence of the external field. Here, we show from numerical experiments that use the material parameters of Mn$_{1.4}$Pt$_{0.9}$Pd$_{0.1}$Sn which are extracted from \cite{JenaElliptical} that an isolated elliptical skyrmion can be nucleated from the stripe domain state or labyrinth domain state by an MFM tip in a systematic way for the system consisting of at least a few atomic layers.
 
The paper is organized as follows: in Section \ref{BPModelElliptical} we discuss elliptical skyrmions in the context of the Belavin-Polyakov (BP) model. In Section \ref{DMIEner} we show why the DMI that supports isotropic skyrmions and antiskyrmions, respectively can also support elliptical antiskyrmions or skyrmions, respectively. In Section \ref{SizeModel} we discuss the size of the elliptical skyrmion. In Section \ref{MagMoment}, we compute the magnetic moment of the elliptical skyrmion and the contribution to the energy due to the lattice-discreteness term. In Section \ref{Creation} we discuss the creation of the elliptical skyrmion using a MFM tip. The results and implications are discussed in Section \ref{Sec_Conclusion} .

\begin{figure}[ht]
\hspace{-0.5cm}
\centering
\includegraphics[width=8cm]{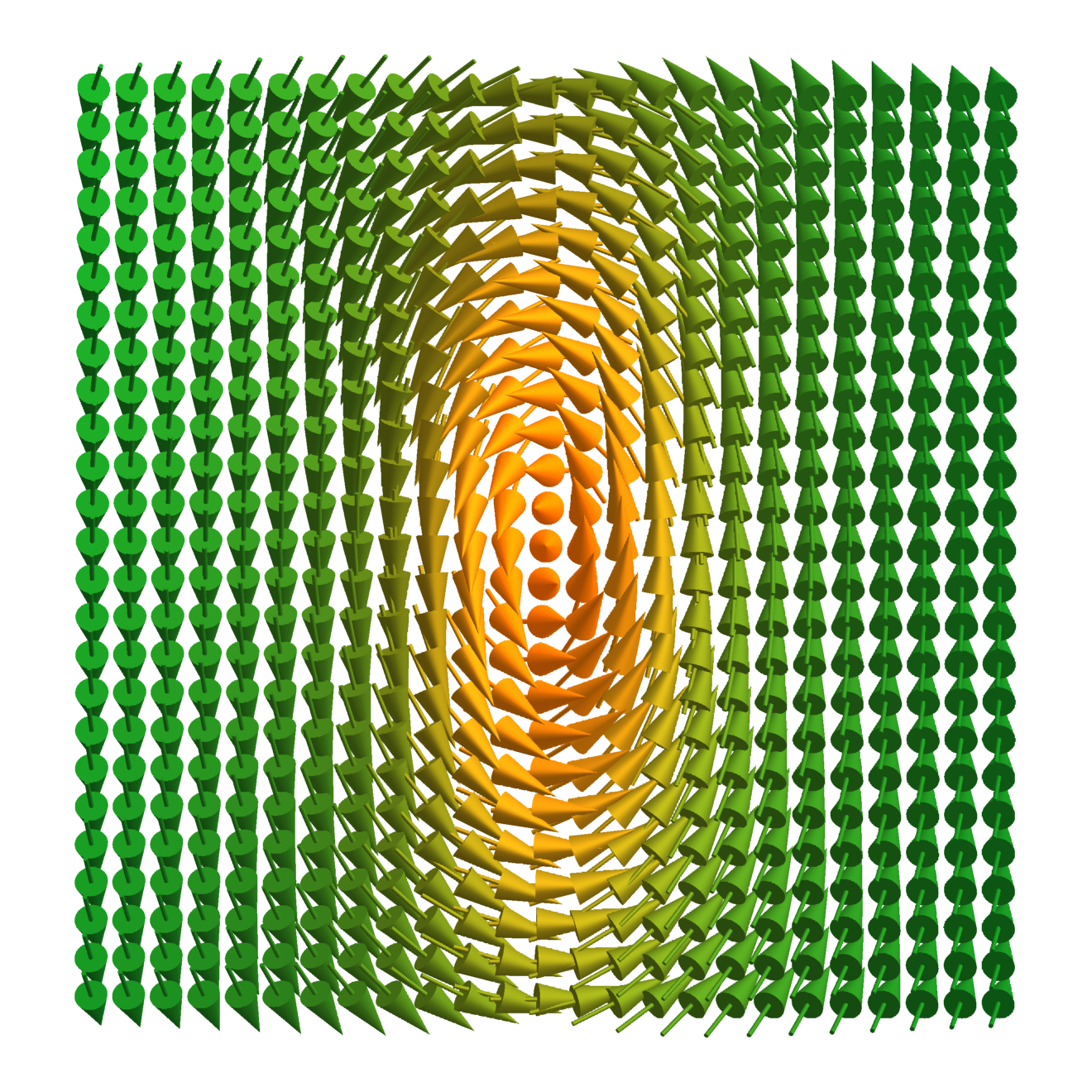}
\caption{An elliptical Bloch skyrmion where the length $\lambda_{y}$ is three times the size of the width $\lambda_{x}$. We take the convention that the skyrmion points up $s_{z}=1$ at its center and down $s_{z}=-1$ far away.}
\label{EllipPic}
\end{figure}

\section{Elliptical Skyrmions: The BP Model}\label{BPModelElliptical}
In the model for the nearest-neighbor Heisenberg exchange interaction for the normalized two-dimensional (2D), three-component spin field $\mathbf{s}^2(x,y)=1$, the exchange energy in the continuous limit can be written as:
\begin{equation}
E_{ex} =\frac{J}{2} \int d^2\rho\left[ \left(\frac{\partial \mathbf{s}}{\partial x}\right)^2+  \left(\frac{\partial \mathbf{s}}{\partial y}\right)^2\right] \label{Exchange},
\end{equation}
for the nearest-neighbor ferromagnetic exchange constant $J>0$. In this model, which we call the pure exchange model, Belavin and Polyakov \cite{BelavinPolyakovMetastable} found that there exist metastable spin configurations that belong to different homotopic classes characterized by an integer called the topological charge. Formally, the topological charge can be obtained from: 
\begin{equation}
Q =\frac{1}{4\pi}\int d^2\rho \quad \mathbf{s} \cdot \left(\frac{\partial \mathbf{s}}{\partial x} \times\frac{\partial \mathbf{s}}{\partial y}  \right) \label{Q}.
\end{equation}
It describes how many times the physical spin vector can be mapped to fully cover the two-sphere as one traverses the entire plane. The metastable spin configurations have a scale-invariant exchange energy proportional to the topological charge: $E_{ex}=4\pi J |Q|$. 

Belavin and Polyakov found that if one maps the spin components to the complex plane via the scalar function\cite{BelavinPolyakovMetastable}:
\begin{equation}
\omega = \frac{s_{x}+i s_{y}}{1-s_{z}} \label{Om}, 
\end{equation}
where $i =\sqrt{-1}$, then $\omega$ yields a metastable spin configuration if it satisfies \cite{BelavinPolyakovMetastable}:
\begin{equation}
\frac{\partial \omega}{\partial x} = \mp i \frac{\partial \omega}{\partial y}, 
\end{equation}
i.e. the Cauchy-Riemann equations that are satisfied by a function that is analytic except for isolated poles. The Belavin-Polyakov (BP) spin profile of the $Q=1$ ($Q=-1$) skyrmion (antiskyrmion) can be found from the relations \cite{CapicBiskyrmionLattices}:
\begin{equation}
\omega_{skyrmion} = \frac{\lambda}{z^{*}}e^{i\gamma}; \quad  \omega_{antiskyrmion}=\frac{\lambda}{z}e^{i\gamma} \label{Omega}.
\end{equation}
Here, $\lambda$ is the size and $\gamma$ determines the orientation of the in-plane spin components, which we call the chirality. The exchange energy is $E_{ex}=4\pi J$ for both the skyrmion and antiskyrmion. 

An elliptical skyrmion breaks the scale invariance of the pure exchange model. This breaking can be introduced into the $\omega$-function by using two parameters:
\begin{equation}
\omega_{ellip}= \frac{e^{i\gamma}}{\frac{x}{\lambda_{x}}- i \frac{y}{\lambda_{y}}}. 
\end{equation}
This is not a minimum energy configuration of the $Q=1$ class. Violating the scale invariance means that the Cauchy-Riemann equations are no longer satisfied:
\begin{equation}
\frac{\partial \omega_{ellip}}{\partial x} \neq \mp i\frac{\partial \omega_{ellip}}{\partial y}.
\end{equation}
Therefore, $E_{ex} >4\pi J $ for the elliptical skyrmion. This can be verified explicitly by obtaining the spin components using the transformations  \cite{CapicBiskyrmionLattices}:
\begin{equation}
s_{z} = \frac{|\omega|^2-1}{|\omega|^2+1}; \quad s_{x}+ i s_{y} = \frac{2\omega}{|\omega|^2+1},
\end{equation}
to obtain:
\begin{eqnarray}
\mathbf{s}_{ellip} &=& \Bigg. \Bigg\{ \frac{2 x \lambda_{x} \lambda_{y}^2 \cos\gamma - 2 y \lambda_{x}^2 \lambda_{y} \sin\gamma}{y^2\lambda_{x}^2+ x^2\lambda_{y}^2+\lambda_{x}^2\lambda_{y}^2}, \nonumber \\&&\frac{2 x \lambda_{x} \lambda_{y}^2 \sin\gamma+2 y \lambda_{x}^2 \lambda_{y} \cos\gamma}{y^2\lambda_{x}^2+ x^2\lambda_{y}^2+\lambda_{x}^2\lambda_{y}^2}, \nonumber \\&& - \frac{y^2\lambda_{x}^2+ x^2\lambda_{y}^2-\lambda_{x}^2\lambda_{y}^2}{y^2\lambda_{x}^2+ x^2\lambda_{y}^2+\lambda_{x}^2\lambda_{y}^2} \Bigg. \Bigg\} \label{SEllip}.
\end{eqnarray}
A picture of an elliptical Bloch skyrmion spin profile is plotted in Fig. \ref{EllipPic}. Then, one obtains the exchange energy for the elliptical skyrmion by substituting Eq. (\ref{SEllip}) into Eq. (\ref{Exchange}) to obtain:
\begin{equation}
E_{ex} =2\pi J \frac{\lambda_{x}}{\lambda_{y}}\left(1+\frac{\lambda_{y}^2}{\lambda_{x}^2} \right).
\end{equation}
This function has a minimum when $\lambda_{x}=\lambda_{y}$ corresponding to the isotropic skyrmion which has the minimum exchange energy of the $Q=1$ homotopic class.  

\section{DMI and Elliptical Skyrmions and Antiskyrmions} \label{DMIEner} 

\begin{figure}[ht]
\hspace{-0.5cm}
\centering
\includegraphics[width=8cm]{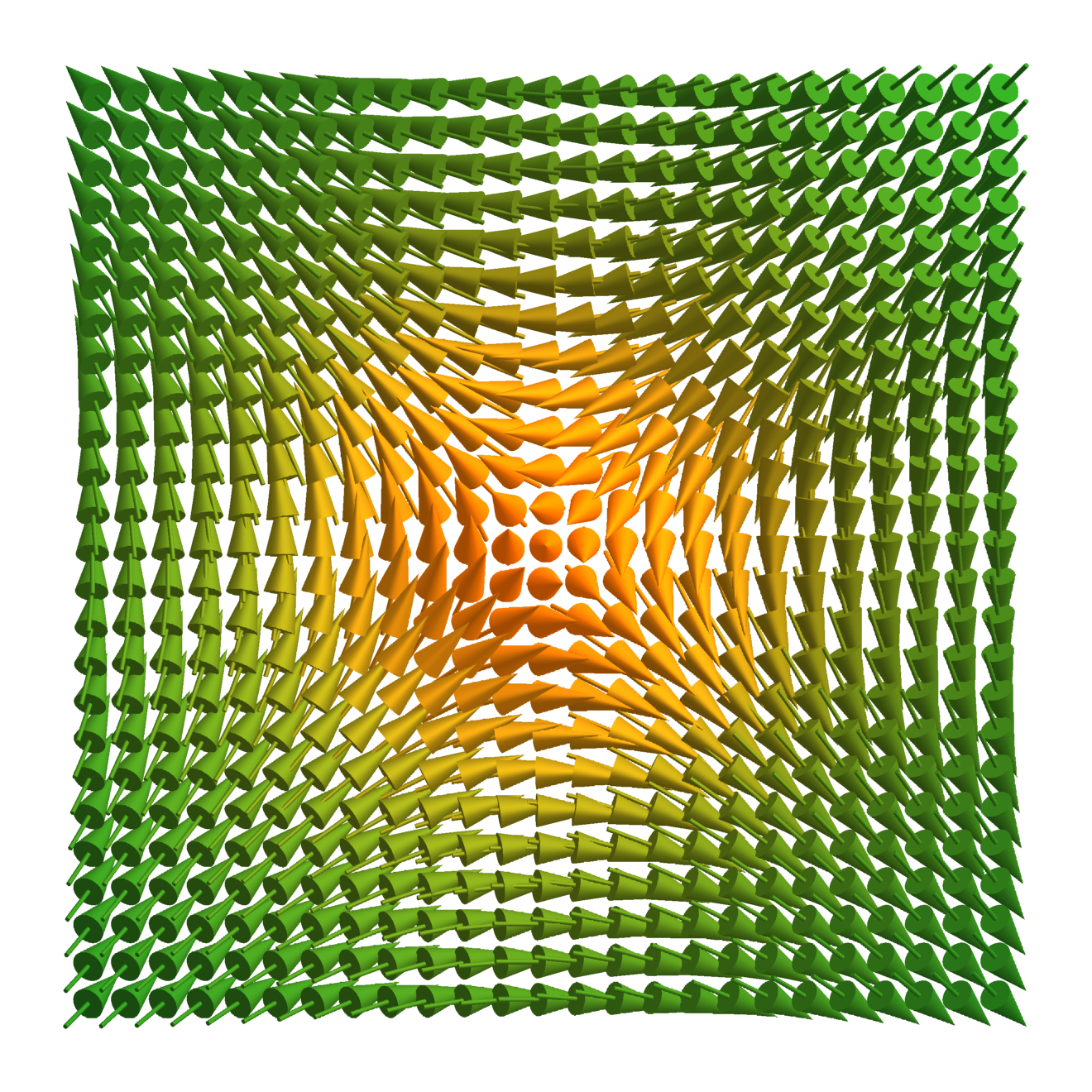}
\caption{Bloch-type antiskyrmion of chirality $\gamma= -\pi/2$. Note that a similar picture of the $\gamma=\pi/2$ Bloch antiskyrmion appeared in \cite{CapicBiskyrmionLattices}. }
\label{Anti}
\end{figure}

In a real system that contains interactions beyond the exchange, such as the DMI, the elliptical skyrmion can be metastable. In the continuous limit, the DMI energy due to the $D_{2d}$ crystal symmetry can be expressed as:
\begin{eqnarray}
E_{DM}&=&-\frac{A}{a}\int d^2\rho \Bigg.\Bigg[s_{z} \times \frac{\partial s_{x}}{\partial y }-s_{x} \times\frac{\partial s_{z}}{\partial y}\nonumber \\&& +s_{z}\times\frac{\partial s_{y}}{\partial x}-s_{y}\frac{\partial s_{z}}{\partial x}\Bigg.\Bigg]. 
\end{eqnarray}
We use the DMI constant $A$ to be consistent with previous works and $a$ is the lattice spacing. We assume that the exchange interaction is much stronger than other interactions, which means that the spin profile is close to the BP shape. Substituting the spin components of the BP isotropic skyrmion by setting $\lambda_{x}=\lambda_{y}$ in Eq. (\ref{SEllip}), one obtains $E_{DM}=0$.  On the other hand, for the BP isotropic antiskyrmion (set $\lambda_{x}=\lambda_{y}$ in Eq. (\ref{NoRepeats})), one obtains $E_{DM} = -\frac{4\pi A}{a} \lambda \sin\gamma$. 
This favors Bloch antiskyrmions of chirality $\gamma=\pi/2$ for $A>0$, which will have in-plane spin components that radiate outwards from the center along the line $y=x$ or  Bloch antiskyrmions of chirality $\gamma=-\pi/2$ for $A<0$ which will have in-plane spin components that radiate inwards from the center along the line $y=x$. The latter is pictured in Fig \ref{Anti} .

Elliptical skyrmions have a non-zero DMI energy. Substituting Eq. (\ref{SEllip}) into the continuous expression for the DMI energy, one obtains:
\begin{equation}
E_{DM} = -\frac{4\pi A (\lambda_{y} -\lambda_{x})\sin\gamma}{a} \label{EllipticalDMIEnergy}. 
\end{equation}
This favors the elongation of the skyrmion along \textit{y} when it is a counterclockwise Bloch skyrmion of chirality $\gamma=\pi/2$ and elongation along \textit{x} when it is a clockwise Bloch skyrmion  of chirality $\gamma=-\pi/2$. However, the elliptical skyrmion approaches being energetically equivalent to the antiskyrmion when the elongation along a given direction is significant. It seems that in a real material such as the one Jena \textit{et al.} \cite{JenaElliptical} studied, elliptical skyrmions must be stabilized by a complicated interplay between the DMI, perpendicular magnetic anisotropy and dipole-dipole interaction. 

As discussed in the introduction, the DMI due to different crystal symmetries favors Bloch or N\'eel-type isotropic skyrmions or antiskyrmions. However, the above calculation suggests that in principle, there can be elongated antiskyrmions in typically skyrmion-hosting materials that have the isotropic Bloch or N\'eel-type DMI characterized by a single DMI constant. To verify this, we obtain the spin components of the elliptical antiskyrmion by defining the complex function $\omega$ as:
\begin{equation}
\omega =\frac{e^{i\gamma}}{\frac{x}{\lambda_{x}}+i \frac{y}{\lambda_{y}}}.
\end{equation}
This yields the spin profile of the $Q=\pm 1$ elliptical skyrmion (antiskyrmion) in the form:
\begin{eqnarray}
\mathbf{s}_{ellip} &=& \Bigg. \Bigg\{ \frac{2 x \lambda_{x} \lambda_{y}^2 \cos\gamma - 2 Q y \lambda_{x}^2 \lambda_{y} \sin\gamma}{y^2\lambda_{x}^2+ x^2\lambda_{y}^2+\lambda_{x}^2\lambda_{y}^2}, \nonumber \\&&\frac{2 x \lambda_{x} \lambda_{y}^2 \sin\gamma+2 Q y \lambda_{x}^2 \lambda_{y} \cos\gamma}{y^2\lambda_{x}^2+ x^2\lambda_{y}^2+\lambda_{x}^2\lambda_{y}^2}, \nonumber \\&& - \frac{y^2\lambda_{x}^2+ x^2\lambda_{y}^2-\lambda_{x}^2\lambda_{y}^2}{y^2\lambda_{x}^2+ x^2\lambda_{y}^2+\lambda_{x}^2\lambda_{y}^2} \Bigg. \Bigg\} \label{NoRepeats}.
\end{eqnarray}

The Bloch DMI in the continuous limit can be expressed as:
\begin{eqnarray}
E_{DM}&=&-\frac{A}{a}\int d^2\rho \Bigg.\Bigg[s_{x} \times \frac{\partial s_{z}}{\partial y }-s_{z} \times\frac{\partial s_{x}}{\partial y}\nonumber \\&& +s_{z}\times\frac{\partial s_{y}}{\partial x}-s_{y}\frac{\partial s_{z}}{\partial x}\Bigg.\Bigg]. 
\end{eqnarray}
If one evaluates the Bloch DMI energy of the isotropic BP antiskyrmion, they will obtain $E_{DM}=0$, while the isotropic Bloch skyrmion has the energy  $E_{DM} = -\frac{4\pi A}{a} \lambda \sin\gamma$. However, the elliptical antiskyrmion has the same energy as Eq. (\ref{EllipticalDMIEnergy}). Thus, it seems that the following is true: materials that support isotropic antiskyrmions can support elliptical skyrmions and vice versa. From the subsequent numerical analysis performed in Section \ref{Creation} , we predict that a typically skyrmion-hosting material where the DMI strength is of the order of the perpendicular magnetic anistropy (PMA) $A a/J \sim D a^2/J$, and the PMA is of the order of the dipole-dipole interaction (DDI), $\beta \sim 1$ (see Section \ref{Creation} ), can support elliptical antiskyrmions. 

\section{Effective Size of the Elliptical Skyrmion} \label{SizeModel}
One can compute the effective size of the elliptical skyrmion by evaluating the following integral:
\begin{equation}
\int d^2 \rho \left(s_{z}+1\right)^{n}.
\end{equation}
This gives the same expression derived by Cai \textit{et al.} for the isotropic BP skyrmion, which in the discrete model can be computed from   \cite{CaiCollapse}:
\begin{equation}
\lambda_{eff,n}^2\equiv \lambda_{x}\lambda_{y}= \frac{n-1}{2^{n} \pi}a^2 \sum_{i} (s_{iz}+1) ^{n} \label{LamEff}
\end{equation}
One can decouple $\lambda_{x}$ and $\lambda_{y}$ in a way similar to the above by going to the continuous limit. If the skyrmion is centered at the coordinate origin, then the size can be found by evaluating:
\begin{eqnarray}
\int_{-\infty}^{\infty} dx (s_{z}(x, 0)+1)^4&=& 5\pi \lambda_{x}\nonumber \\
\int_{-\infty}^{\infty} dy (s_{z}(0, y)+1)^4&=& 5\pi \lambda_{y},
\end{eqnarray}
so that the size for the discrete case can be found from the sums:
\begin{eqnarray}
\lambda_{eff,x} &=&  \frac{1}{5\pi} \sum_{i=1}^{N_{x}} (\mathbf{s}(n_{ix}, y_{pos,skyrm})_{z}+1)^4 \nonumber \\
\lambda_{eff,y} &=&  \frac{1}{5\pi} \sum_{i=1}^{N_{y}} (\mathbf{s}( x_{pos,skyrm}, n_{iy})_{z}+1)^4 \label{LamEffx}. 
\end{eqnarray}
Then, the above will give the width and length of the elliptical skyrmion. One can compare the accuracy of these individual numbers to the area of the skyrmion computed via Eq. (\ref{LamEff}), where one expects $\lambda_{x}\lambda_{y}= \lambda^2$. The results are in agreement, provided one can accurately identify the center of the elliptical skyrmion. The center can be computed numerically by finding the location on the lattice where $s_{z}$ is a maximum. In principle however, the skyrmion center can be located in between lattice spacings, a location that is ill-defined in the discrete numerical program, and lead to some disagreement. 

\section{Magnetic Moment and Lattice Energy of the Elliptical Skyrmion}  \label{MagMoment}
The magnetic moment of the elliptical skyrmion can be computed from:
\begin{eqnarray}
S_{z} &=& \frac{1}{a^2} \int d^2\rho \left(s_{z}+1\right)  \\&=&  \frac{1}{a^2} \int \rho d\rho d\phi \frac{2  \lambda_{x}^2 \lambda_{y}^2}{\rho^2\lambda_{y}^2 \cos^2\phi+ \rho^2 \lambda_{x}^2 \sin^2\phi+ \lambda_{x}^2\lambda_{y}^2},\nonumber
\end{eqnarray}
where we subtract the contribution from the uniform ferromagnetic background. After performing the $\phi$ integration:
\begin{equation}
S_{z} = \frac{4\pi \lambda_{x}\lambda_{y}}{a^2} \int d\rho \frac{\rho}{\sqrt{(\rho^2+\lambda_{x}^2)(\rho^2+\lambda_{y}^2)}}.
\end{equation}
Recognizing that this diverges for the infinite system, we integrate over the system size $L$:
\begin{eqnarray}
S_{z} &=& \frac{4\pi \lambda_{x}\lambda_{y}}{a^2} \int_{0}^{L} d\rho \frac{\rho}{\sqrt{(\rho^2+\lambda_{x}^2)(\rho^2+\lambda_{y}^2)}}\nonumber \\&=&\frac{4\pi \lambda_{x}\lambda_{y}}{a^2}\; \textrm{ln}\left(\frac{\sqrt{L^2 +\lambda_{x}^2}+ \sqrt{L^2+\lambda_{y}^2}}{\lambda_{x}+\lambda_{y}} \right).
\end{eqnarray}
When the system size is much larger than the skyrmion size,  $L \gg \lambda_{x}, \lambda_{y}$, this reduces to:
\begin{equation}
S_{z}=\frac{4\pi \lambda_{x}\lambda_{y}}{a^2}\; \textrm{ln}\left(\frac{2L}{\lambda_{x}+\lambda_{y}} \right). 
\end{equation}
Thus, the magnetic moment of the elliptical skyrmion is proportional to its area, and very similar to the magnetic moment of the isotropic skyrmion computed in \cite{ChoukQuantum}. In principle, one can use the disparate lengths $\lambda_{x}$ and $\lambda_{y}$ as additional degrees of freedom for potential computing applications involving skyrmions.

One can also compute the energy of the elliptical skyrmion due to the discreteness of the lattice by evaluating the term \cite{CaiCollapse}: 
\begin{equation}
E_{lat} = -\frac{J a^2}{24} \int d^2\rho \left\{\left(\frac{\partial^2\mathbf{s}}{\partial x^2}\right)^2+\left(\frac{\partial^2\mathbf{s}}{\partial y^2}\right)^2 \right\},
\end{equation}
which is the next higher order term in the exchange energy if one does a Taylor series expansion of the nearest-neighbor spins. Upon substitution of Eq. (\ref{SEllip}), one obtains:
\begin{equation}
E_{lat} = - \frac{\pi J a^2 (\lambda_{x}^4 +\lambda_{y}^4)}{3 \lambda_{x}^3\lambda_{y}^3}. 
\end{equation}
This term promotes the collapse of the elliptical skyrmion in the pure exchange model and converges to the result \cite{CaiCollapse} $E_{lat} = -\frac{2\pi J a^2}{3\lambda^2}$ for the isotropic case.

\section{Elliptical Skyrmion Creation}\label{Creation}
In the numerical work, we consider a lattice model for a 2D ferromagnetic film  with the energy:
\begin{eqnarray}
{\cal H} & = &- \frac{1}{2} \sum_{i j}J_{ij}\mathbf{s}_{i}\cdot \mathbf{s}_{j}-H_{z}\sum_{i }  s_{iz}\nonumber \\
& - &A \sum_{i} (\mathbf{s}_{i}\times \mathbf{s}_{i+\delta_{y}})_{y}-(\mathbf{s}_{i}\times \mathbf{s}_{i+\delta_{x}})_{x}\nonumber \\ &-& \frac{D}{2}  \sum_{i} s_{iz}^2- \frac{E_{D}}{2}\sum_{ij}\Phi_{ij,\alpha\beta}s_{i\alpha}s_{j\beta}, \label{Hamiltonian}
\end{eqnarray}
which contains the exchange interaction with coupling $J$, the Zeeman interaction from an applied field $H$ directed along the negative \textit{z}-axis, the DMI with strength $A$, the easy axis perpendicular magnetic anisotropy (PMA) with constant $D$ and the dipole-dipole interaction (DDI) characterized by the strength $E_{D}$. We consider the exchange interaction as being for the nearest-neighbors on a square lattice with the same exchange constant $J$ for all pairs. The magnetic field is applied along the negative \textit{z}-axis because we take the convention that the skyrmion's spins point up at its center. The subscripts $\delta_{x}$ and $\delta_{y}$ for the DMI term refer to the next nearest lattice site in the positive $x$ or $y$ direction.  The DMI term favors Bloch-type elliptical skyrmions with chirality $\gamma=\pi/2$ or $\gamma=-\pi/2$ elongated along the \textit{y} or \textit{x}-axis, respectively when $A>0$. 

The DDI term contains:
\begin{equation}
\Phi_{ij,\alpha\beta}\equiv a^{3}r_{ij}^{-5}\left(3r_{ij,\alpha}r_{ij,\beta}-\delta_{\alpha\beta}r_{ij}^{2}\right),\label{DDIPhi}
\end{equation}
where $\mathbf{r}_{ij}\equiv\mathbf{r}_{i}-\mathbf{r}_{j}$ is the 
displacement vector between the lattice sites and $\alpha,\beta=x,y,z$ gives the Cartesian coordinates for the most general case where one considers a 3D model. The relative strength of the PMA to the DDI is given by the dimensionless parameter $\beta \equiv D/(4\pi E_{D})$.  

The effective 2D DDI model that we use for the film of $N_z$ layers assumes that the spin field does not depend on \textit{z}. As a consequence, we treat the effective DDI as  the interaction between columns of parallel spins, in accordance with the methodology first introduced in Ref. [\onlinecite {Capic2019}]. Therefore, when the film has $N_z>1$ layers, Eq. (\ref{Hamiltonian}) gives the energy per layer. At large distances, the effective DDI divided by the number of layers is $N_z \Phi$, where $\Phi$ is defined in Eq. (\ref{DDIPhi}). 

This Hamiltonian contains all relevant interactions that would be present in a real material. We use the parameters of the compound Mn$_{1.4}$Pt$_{0.9}$Pd$_{0.1}$Sn that has the DMI due to the $D_{2d}$ crystal symmetry.  For the lattice spacing $a$ and the exchange constant $J$ we chose $a = J = 1$ in the numerical computations. Following the treatment in Ref. [ \onlinecite{JenaElliptical} ], we obtain the parameters $A/J = 0.0625$, $D/J = 0.04167$ and $\beta = 1.6074$.  The system was studied on a square lattice of $500 \times 500$ spins for systems with $N_z=1$, $N_z=5$, $N_z=10$, $N_z=20$ and $N_z=50$ layers.

The numerical routine used in this work, which is described in [ \onlinecite{Garanin2013} ] involves the minimization of the energy via the alignment of the spins with the effective field $\mathbf{H}_{\mathrm{eff,i}} =-\delta {\cal H}/\delta \mathbf{s}_{i}$ with the probability $\alpha$. Otherwise, the  spins are rotated about the effective field, $\mathbf{s}_{i} \rightarrow 2(\mathbf{s}_{i} \cdot \mathbf{H}_{\mathrm{eff}, i})\mathbf{H}_{\mathrm{eff}, i}/H_{\mathrm{eff}, i}^2 -\mathbf{s}_{i}$ in a process that conserves the energy and allows the system to explore the phase space with the probability $1-\alpha$. Thus, $\alpha$ serves as a relaxation parameter that we take as $\alpha=0.03$.

As a preliminary step, we confirm that the elliptical skyrmion is a metastable state for the parameters of this system. To verify this, we start with the initial state of the isotropic Bloch skyrmion, $\lambda_{x}=\lambda_{y}$ and run the numerical routine to compute the minimum energy. We find, much like Jena \textit{et al.} \onlinecite{JenaElliptical}, that after the relaxation the Bloch skyrmion elongates along the \textit{y}-axis, i.e. $\lambda_{y}>\lambda_{x}$ when the initial state has the chirality $\gamma=\pi/2$ corresponding to a counterclockwise Bloch skyrmion, or the Bloch skyrmion elongates along the \textit{x}-axis, i.e. $\lambda_{x}>\lambda_{y}$ when the initial state has the chirality $\gamma= -\pi/2$ corresponding to the clockwise Bloch skyrmion. This is consistent with the result from  Eq. (\ref{EllipticalDMIEnergy}) for the DMI energy in the continuous limit using the explicit spin components of the elliptical skyrmion from Eq. (\ref{SEllip}).  

Next, we started with the initial spin configuration of a Bloch stripe domain state, see for instance Eq. (11) in [\onlinecite{EzawaCompact}] for the explicit spin components. The initial stripe domain state was allowed to relax in the absence of the magnetic field following the numerical routine described above. This yielded a collection of stripe domains with a width of the order of the separation between them. Next, we applied a magnetic field to this state. If the applied magnetic field was strong enough to induce the ferromagnetic state, as we monitored the system in the course of the computation of the minimum energy, the stripe domains separated from the boundary. This led to an abrupt jump in the topological charge from $Q=0$ to $Q=n$, via the formation of elliptical skyrmions, where $n$ was the initial number of stripe domains. We computed $Q$ using the discretized version of Eq. (\ref{Q}). However, by the end of the numerical computation, the elliptical skyrmions had disappeared, so that the final configuration was the ferromagnetic state.

This suggested it was possible to nucleate elliptical skyrmions from the stripe domain state if the applied magnetic field was stronger at the edges of the film compared to the center. Motivated by this, we were able to create an elliptical skyrmion from the stripe domain state by applying a constant field $H_{z}$ in addition to a gradient field along \textit{z} which had the form:
\begin{equation}
H_{grad, z} = H_{g} \left| 2 \frac{n_{y}}{N_{y}}-1 \right|. \label{Grad} 
\end{equation}

These preliminary findings suggest an elliptical skyrmion could be created with a magnetic force microscope, where the \textit{z}-component of the tip field can locally vary the applied magnetic field. The tip field for the MFM  is modeled as a point magnetic dipole with dipole moment $\mathbf{m}= m \hat{z}$ held at position $r= (x_{0}, y_{0}, - h)$ below the lattice. It has the form:
\begin{eqnarray}
\mathbf{B}_{tip}&=&  \frac{B_{h} h^3}{2} \Bigg.\Bigg\{\frac{ 3 h (x-x_{0})}{((x-x_{0})^2+ (y-y_{0})^2+h^2)^{5/2}},\nonumber\\&& \frac{ 3 h (y-y_{0})}{((x-x_{0})^2+ (y-y_{0})^2+h^2)^{5/2}}, \nonumber\\ &&\frac{ 2h^2-(x-x_{0})^2-(y-y_{0})^2}{((x-x_{0})^2+( y-y_{0})^2+h^2)^{5/2}}\Bigg.\Bigg\}\label{TipField}. 
\end{eqnarray}
Here, the parameter $B_{h}$ is defined so that the ratio $B_{h}/J$ is dimensionless. It is given by \cite{GaraninMFM}:
\begin{equation}
B_{h}=\frac{ g S \mu_{0} \mu_{B} m }{2 \pi h^3} \label{BhDef}
\end{equation}
so that one recovers the usual prefactor $\mu_{0} m/4\pi$. Note that the additional factors $g S \mu_{B}$ come from the Zeeman energy for the total atomic spin $S$ and are also absorbed into $B_{h}$ for convenience. The same is true for our applied field $H_{z}$. The system of $N_{z}>1$ layers  that are stacked on top of each other along increasing \textit{z} has the vertical distance of the tip to any layer $n_{z}$ corresponding to the substitution $h \rightarrow h +n_{z}-1$ into Eq. (\ref{TipField}).

We found that elliptical skyrmions could indeed be nucleated by a magnetic tip for the system of $N_{z}\geq 5$ atomic layers. We started with the initial state of the relaxed Bloch stripe domains in the absence of the magnetic field, $H_{z}=0$. Then, we applied a  magnetic field to the system that was strong enough to induce the ferromagnetic state while the tip was held over the center of one of the stripe domains, providing a locally varying magnetic field that was in the opposite direction of the external magnetic field. For the numerical routine, the tip field enters the energy in Eq. \ref{Hamiltonian}) as an additional Zeeman term of the form $ -\sum_{i} \mathbf{B}_{tip} \cdot \mathbf{s}_{i}$. 

\begin{figure}[ht]
\hspace{-0.5cm}
\centering
\includegraphics[width=8cm]{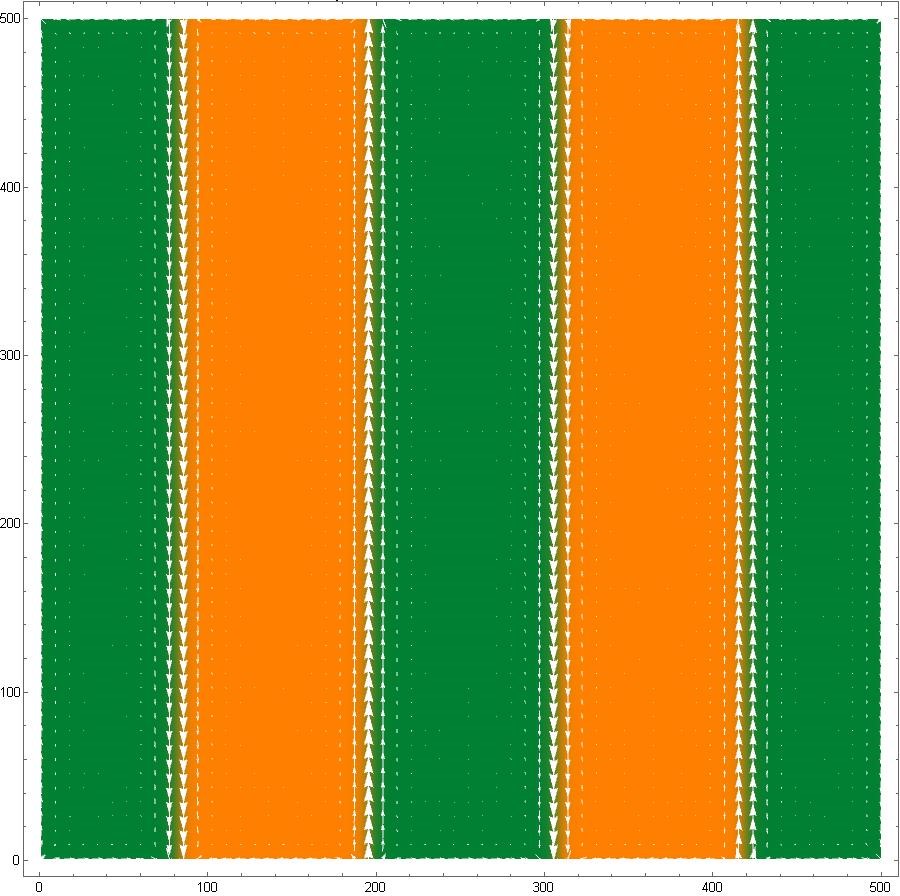}
\includegraphics[width=8cm]{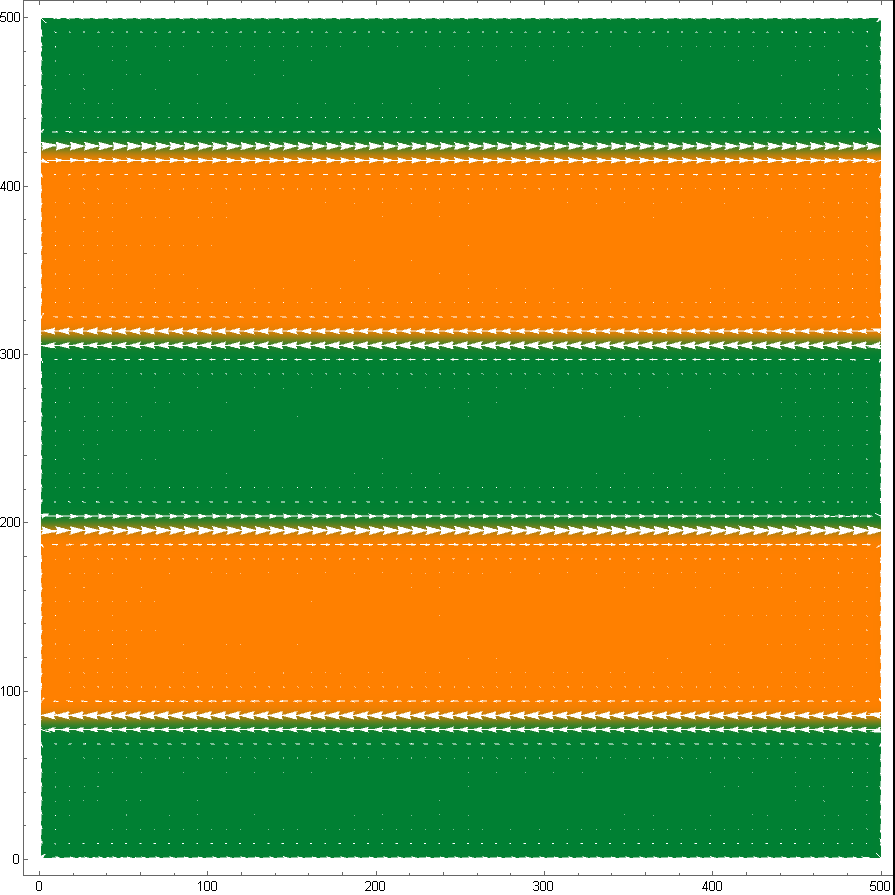}
\caption{These are vertical and horizontal Bloch stripe domains that are used as the initial state. They are found by the numerical computation of the minimum energy in the absence of the magnetic field, $H_{z}=0$. The orange regions indicate where the spins point up, $s_{z}=1$, while the green regions are where the spins point down, $s_{z} = -1$. The in-plane spin components appear as white arrows. }
\label{StripeDomains}
\end{figure}

\begin{figure}[ht]
\hspace{-0.5cm}
\centering
\includegraphics[width=8cm]{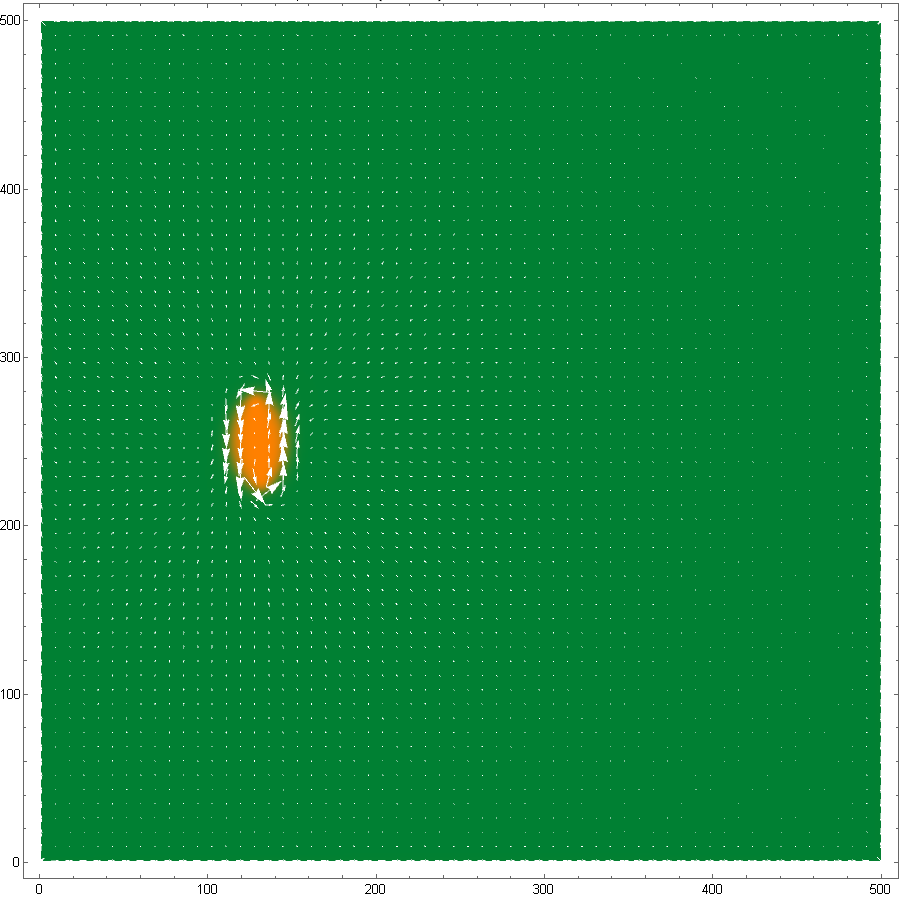}
\includegraphics[width=8cm]{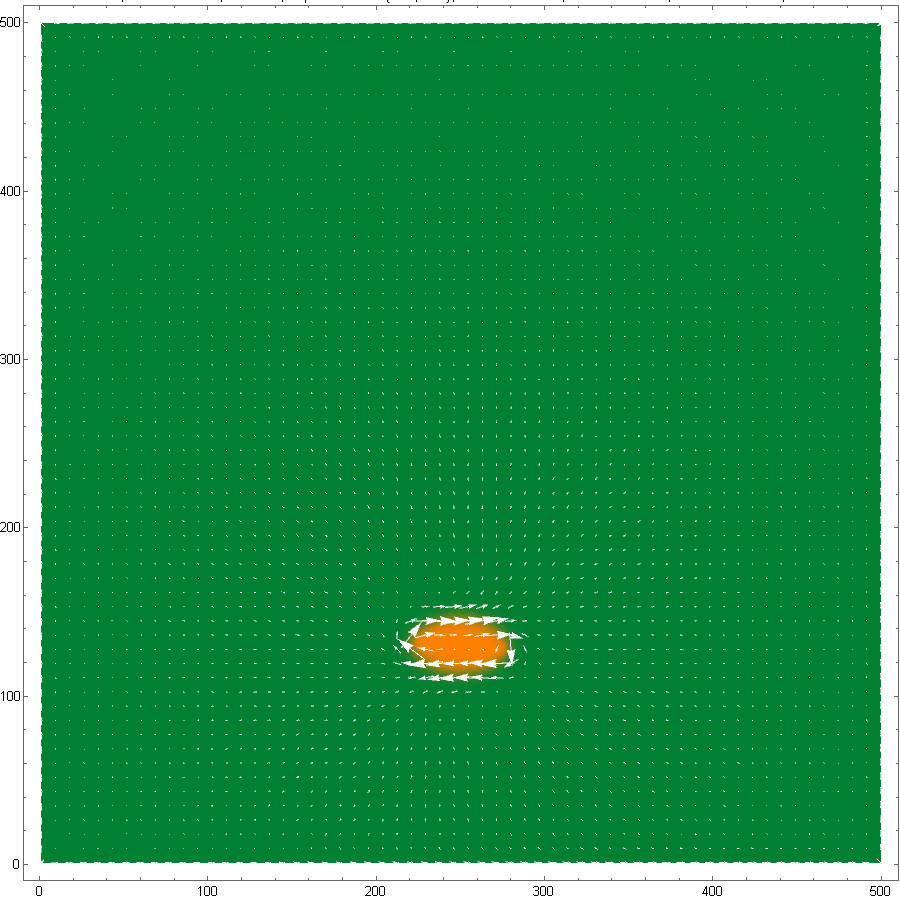}
\caption{Creation of the elliptical skyrmion when the magnetic tip is held at height $h/a=150$ for the system of $N_{z}=50$ layers when $H_{z}/J = -0.017$. The tip is held at position $(n_{x}/a=125, n_{y}/a=250)$ and $(n_{x}/a=250, n_{y}/a=125)$ for the system consisting of two vertical or two horizontal stripe domains pictured in the top and bottom panels of Fig. \ref{StripeDomains}, respectively.}
\label{MFMFig}
\end{figure}

Elliptical skyrmions nucleated by the tip have a chirality determined by the orientation of the initial Bloch stripe domain state. Counterclockwise Bloch skyrmions of chirality $\gamma=\pi/2$ are nucleated from the stripe domain state involving the rotations of the \textit{y} and \textit{z}-components of the spins as one moves along the \textit{x}-axis, i.e. "vertical" stripe domains, while clockwise Bloch skyrmions of chirality $\gamma=-\pi/2$ are nucleated from the stripe domain state involving the rotations of the \textit{x} and \textit{z}-components of the spins as one moves along the \textit{y}-axis, i.e. "horizontal" stripe domains, see the first and second panels of Fig. \ref{StripeDomains}, respectively.
Successful creation of the skyrmion from the vertical and horizontal stripe domain state of Fig. \ref{StripeDomains} is shown in the first and second panels of Fig. \ref{MFMFig}, respectively. 

In addition to the two stripe domain state, we found that the elliptical skyrmion could be nucleated from the single domain state as well. Naively, one would expect in the presence of multiple stripe domains, a magnetic tip placed in between any pair of stripe domains would nucleate a pair or more of elliptical skyrmions. Instead, we find this is not the case. This is due to the competition between the magnetic field produced by the tip and the applied magnetic field. In order to affect multiple domains, the MFM must be held at a larger height below the film to increase its region of influence. In turn, this decreases the magnitude of the magnetic field at the edge of the film, with the net result that the domains will not "peel" away from the edges to form elliptical skyrmions. Consequently, for other states, such as a state consisting of four stripe domains, an elliptical skyrmion can only be nucleated if the tip is held over the center of one of the domains.

\begin{figure}[ht]
\hspace{-0.5cm}
\centering
\includegraphics[width=8cm]{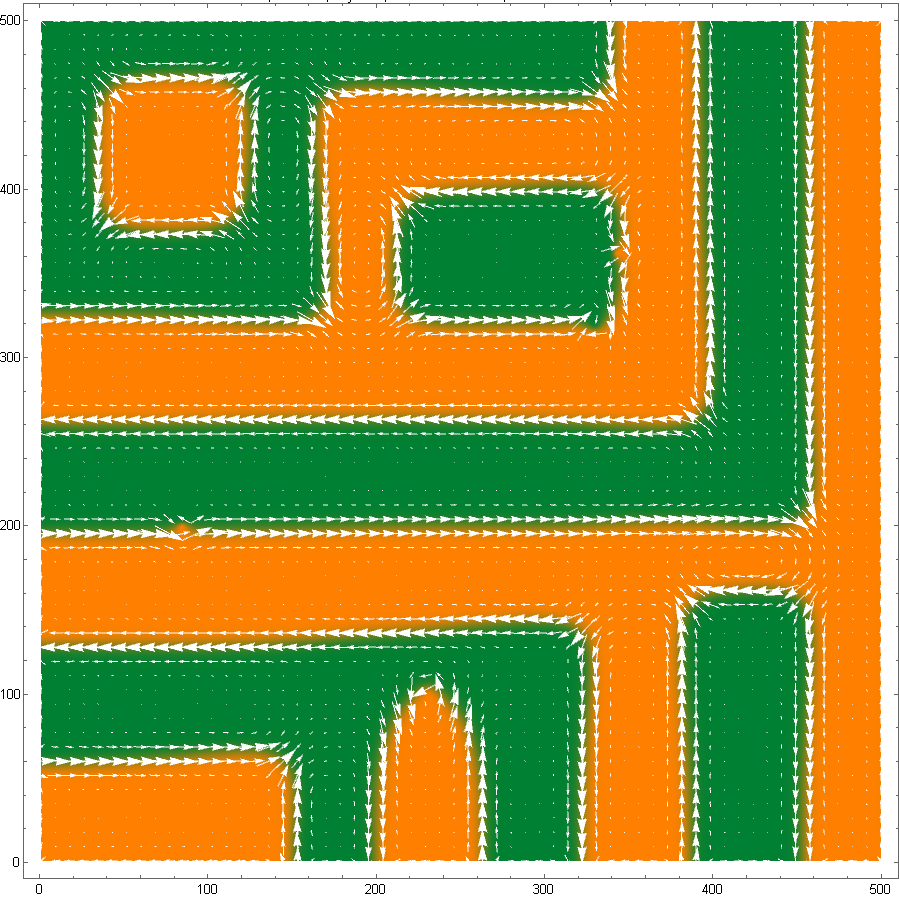}
\includegraphics[width=8cm]{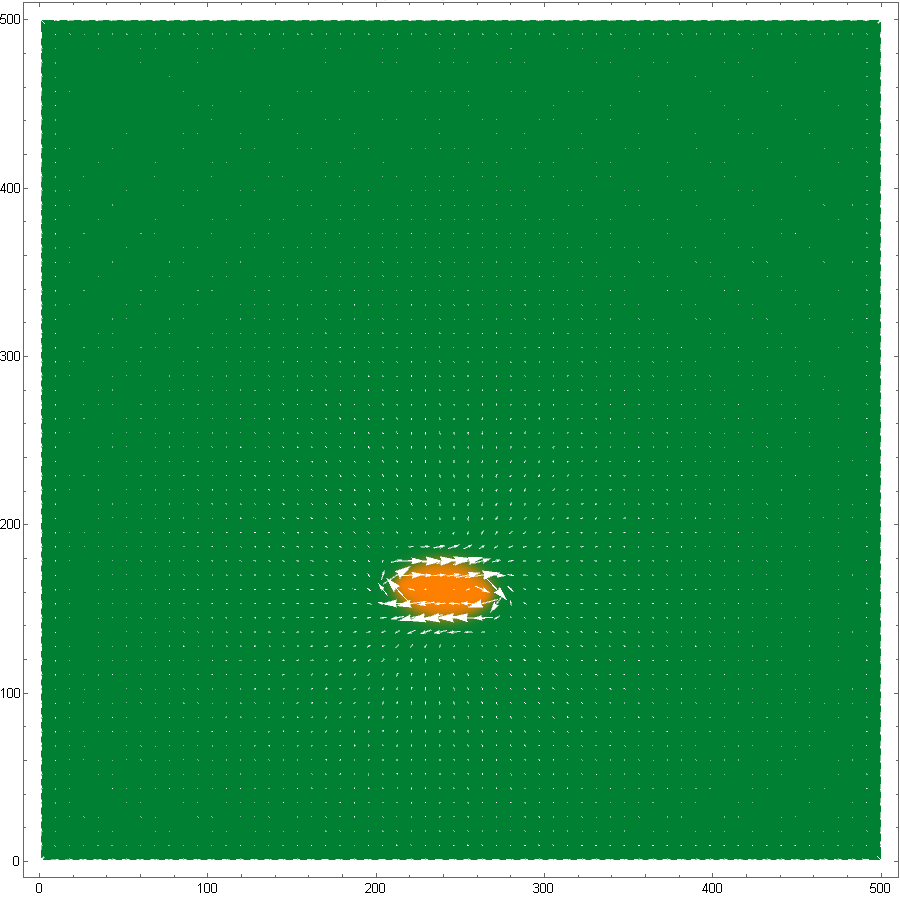}
\caption{First panel: Labyrinth domain state for the system of $N_{z}=50$ layers. Elliptical skyrmions appear as connections that arise when the stripe domains intersect.\\ Second panel: A horizontal elliptical skyrmion nucleated when the tip is held at a position $\{240, 160\}$ at a height of $h/a=100$ for $H_{z}/J = -0.017$.}
\label{Lab}
\end{figure}

\begin{figure}[ht]
\hspace{-0.5cm}
\centering
\includegraphics[width=8cm]{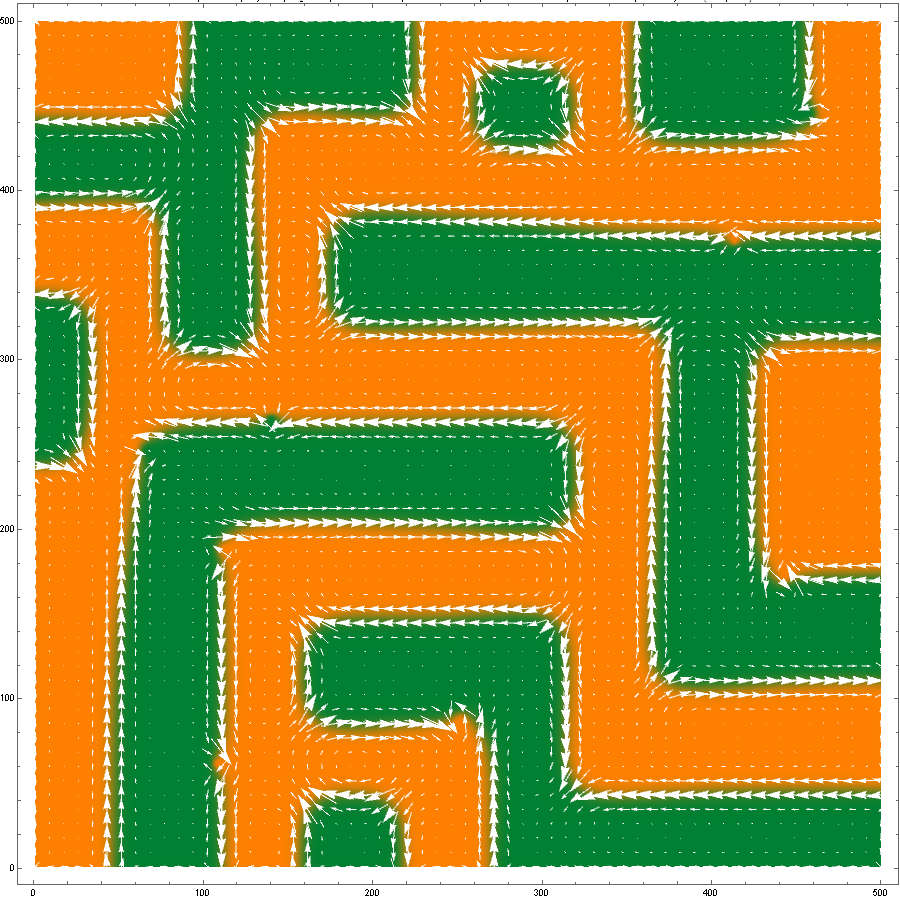}
\includegraphics[width=8cm]{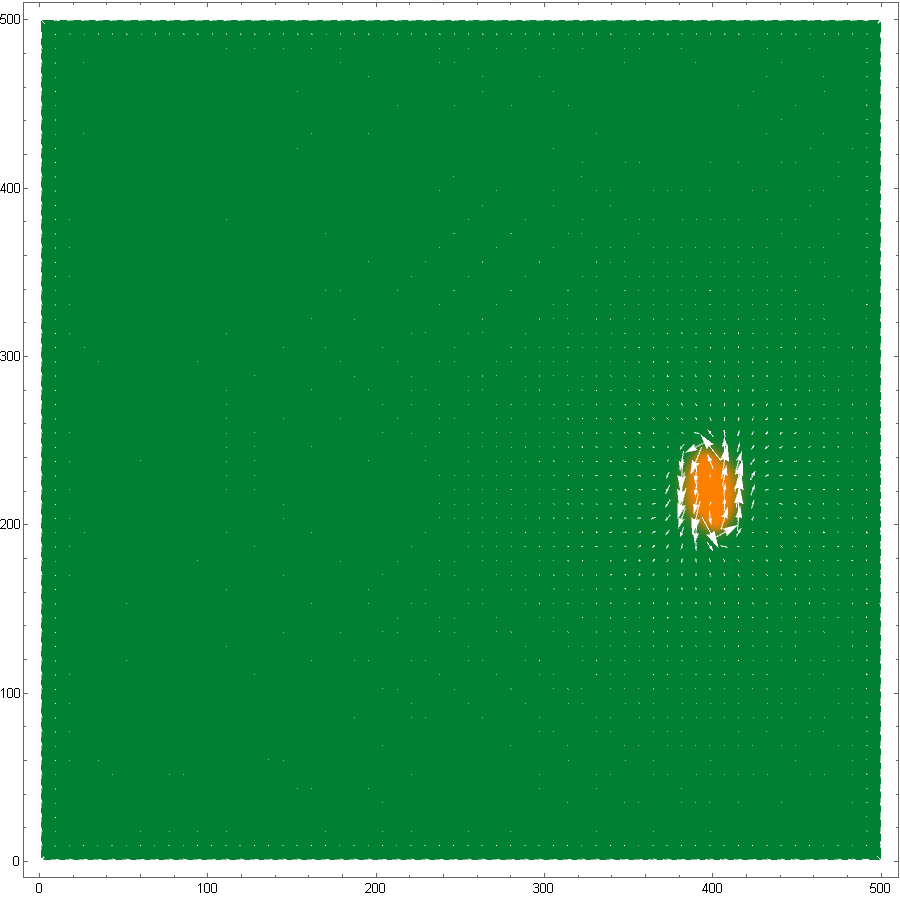}
\caption{First panel: Another labyrinth domain state for the system of $N_{z}=50$ layers. Elliptical skyrmions appear as connections that arise when the stripe domains intersect. Second panel: A vertical elliptical skyrmion nucleated when the tip is held at a position $\{400, 220\}$ at a height of $h/a=50$ for the system of $N_{z}=50$ layers and $H_{z}/J = -0.017$.}
\label{Lab2}
\end{figure}

Additionally, we found that an elliptical skyrmion could be nucleated from a labyrinth domain state if placed at the proper location and held relatively close to the film. The labyrinth domain spin configuration was obtained by starting with the initial state consisting of a random alignment of spins and allowing the system to relax in the absence of an applied magnetic field following the routine that involved the numerical minimization of the energy. One such final configuration is shown in the first panel of Fig. \ref{Lab}. When a magnetic tip is held below the film and an external field is applied, a horizontal elliptical skyrmion can be written from the labyrinth domain when the tip is placed near a section of stripe domain with a horizontal orientation, see the second panel of Fig. \ref{Lab}.
In another labyrinth domain configuration, a vertical elliptical skyrmion was created when the tip was placed near a section of stripe domain with a vertical orientation, see Fig. \ref{Lab2}. For reasons that will be discussed shortly, we were able to write the elliptical skyrmion from the labyrinth domain for all cases except $N_{z}=5$ when the nucleation of the skyrmion becomes very difficult.

We wanted to study the nucleation process more systematically by attempting to determine the minimum tip field strength $B_{h}$ needed to nucleate the elliptical skyrmion as a function of the height $h$ at which the magnetic tip is held below the film.  In an actual experiment, the dipole tip strength would be fixed and instead one would vary the height that it is held below the film, but it is easier to vary the field strength for fixed tip height in numerical experiment. It follows from the definition of $B_{h}$ from Eq. (\ref{BhDef}) that the quantity $B_{h} h^3 \propto m$, the dipole moment of the magnetic tip. 

\begin{figure}[ht]
\hspace{-0.5cm}
\centering
\includegraphics[width=8cm]{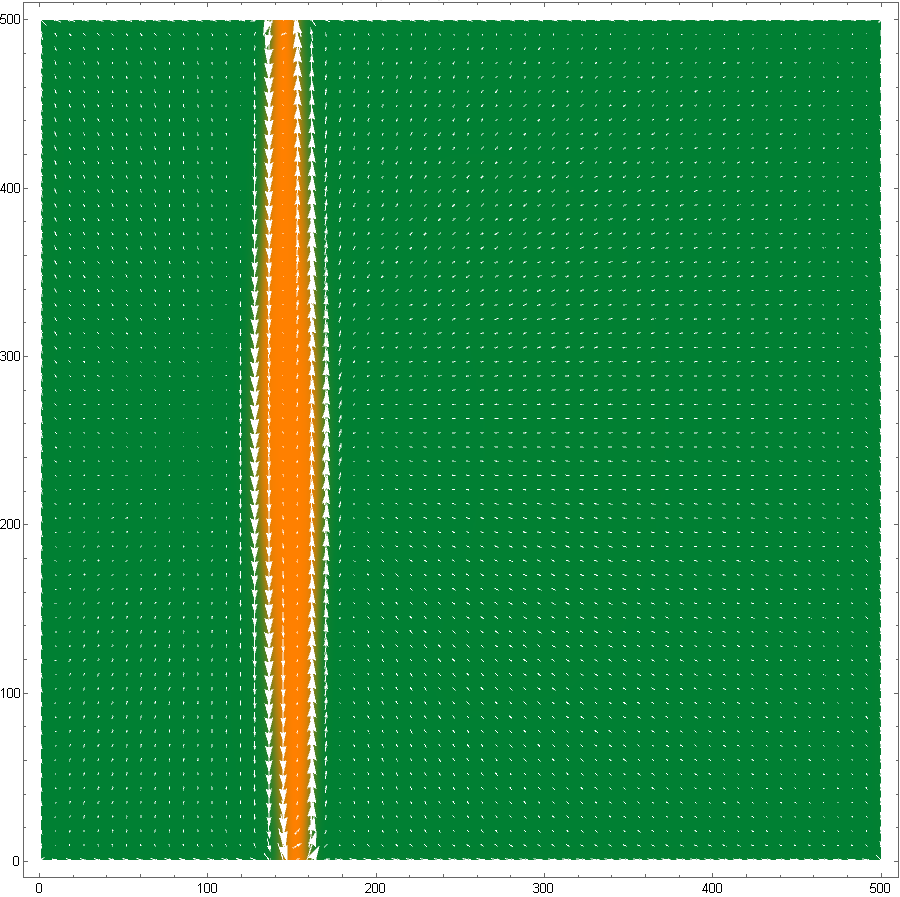}
\caption{Bulging of the stripe domain when the magnetic tip is held too far below the film. }
\label{Bulge}
\end{figure}

Generally speaking, we found that a smaller tip field is required to nucleate the elliptical skyrmion when the magnetic tip is held further below the film. Additionally, the applied field must be strong enough to annihilate the stripe domain state in the absence of the magnetic tip. However, if the tip is held too far below the film, there reaches a point where rather than nucleating the skyrmion from the stripe domain, it will simply cause the stripe domain to bulge near the location of the tip, see Fig. \ref{Bulge}. This regime occurs when $h\sim N/2$ for the square lattice of length $N$. If the skyrmion is nucleated, the qualitative behavior of the tip field is roughly $B_{h}(h)\sim 1/\sqrt{h}$, within a certain height regime. For instance, when $N_{z}=50$, $B_{h}\propto h^{-0.46}$ for larger tip heights $h\geq 100$ and becomes progressively steeper with decreasing $h$.

To begin, we studied the affect of changing the applied magnetic field on the nucleation process for a system with a fixed number of atomic layers $N_{z}=50$. Without loss of generality, we used the starting configuration of the two vertical stripe domain state from the first panel of Fig. \ref{StripeDomains}.  The dipole strength was decreased in small increments with a maximum step size of $dB_{h}/J=-0.0005$ until the elliptical skyrmion collapsed. The value before the skyrmion collapse was recorded as our $B_{h}$. The step size was sufficiently small enough that the error bars are smaller than the points plotted on any given graph. 

\begin{figure}[ht]
\hspace{-0.5cm}
\centering
\includegraphics[width=8cm]{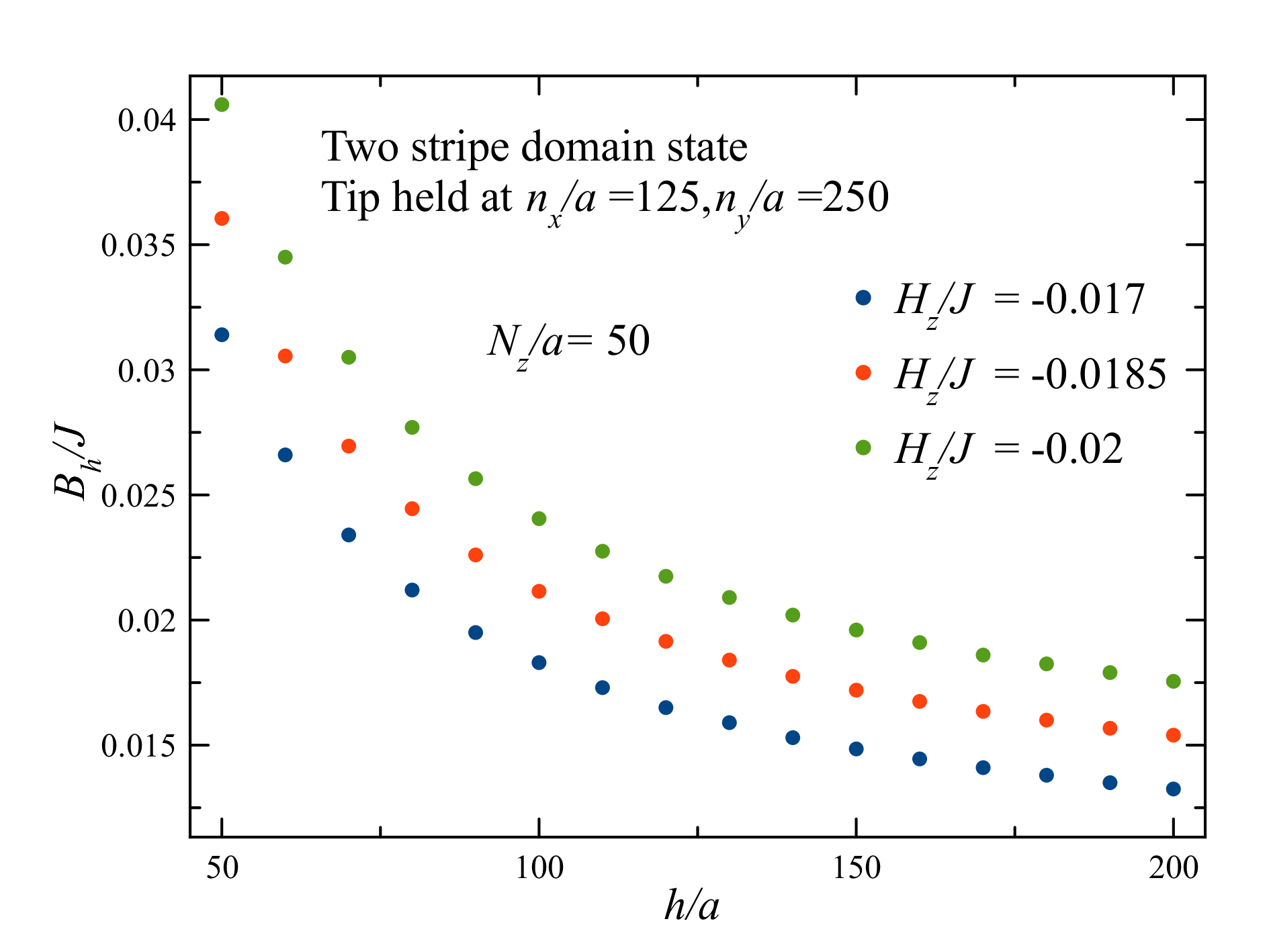}
\caption{Minimum tip field strength $B_{h}/J$ needed to nucleate the elliptical skyrmion as a function of tip height for fixed atomic layer number $N_{z}=50$. }
\label{TipHz}
\end{figure}

We found that increasing the magnitude of the applied magnetic field $H_{z}$ did not change the behavior of the minimum dipole strength needed to nucleate the elliptical skyrmion as a function of the tip height $h$, see Fig. \ref{TipHz}. Using the data extracted from Fig. \ref{TipHz}, one finds that the required dipole moment of the MFM needed to nucleate the skyrmion follows a power law given by $m \propto h^{2.56}$ independent of $H_{z}$. The prefactors,  listed in order of increasing magnitude $H_{z}$  are 0.132, 0.162 and 0.178. We conclude from this that the prefactors contain the dependence of the minimum dipole strength needed to nucleate the skyrmion on the external field. The fit is done to the independent variable $h$ for comparison purposes used later on to get an indication of the relative strengths of the required dipole moment for nucleation. Furthermore, the minimum dipole moment needed to nucleate the skyrmion as a function of the tip field strength has a more complicated behavior, as the minimum tip field is a more complicated function that also depends on the tip height and the applied field. Similarly, a plot of $B_{h}$ vs $1/h^3$ (see Eq. \ref{BhDef}) is non-linear at sufficiently small $h$. In contrast, the fits $m(h)$ follow a smooth power law.

\begin{figure}[ht]
\hspace{-0.5cm}
\centering
\includegraphics[width=8cm]{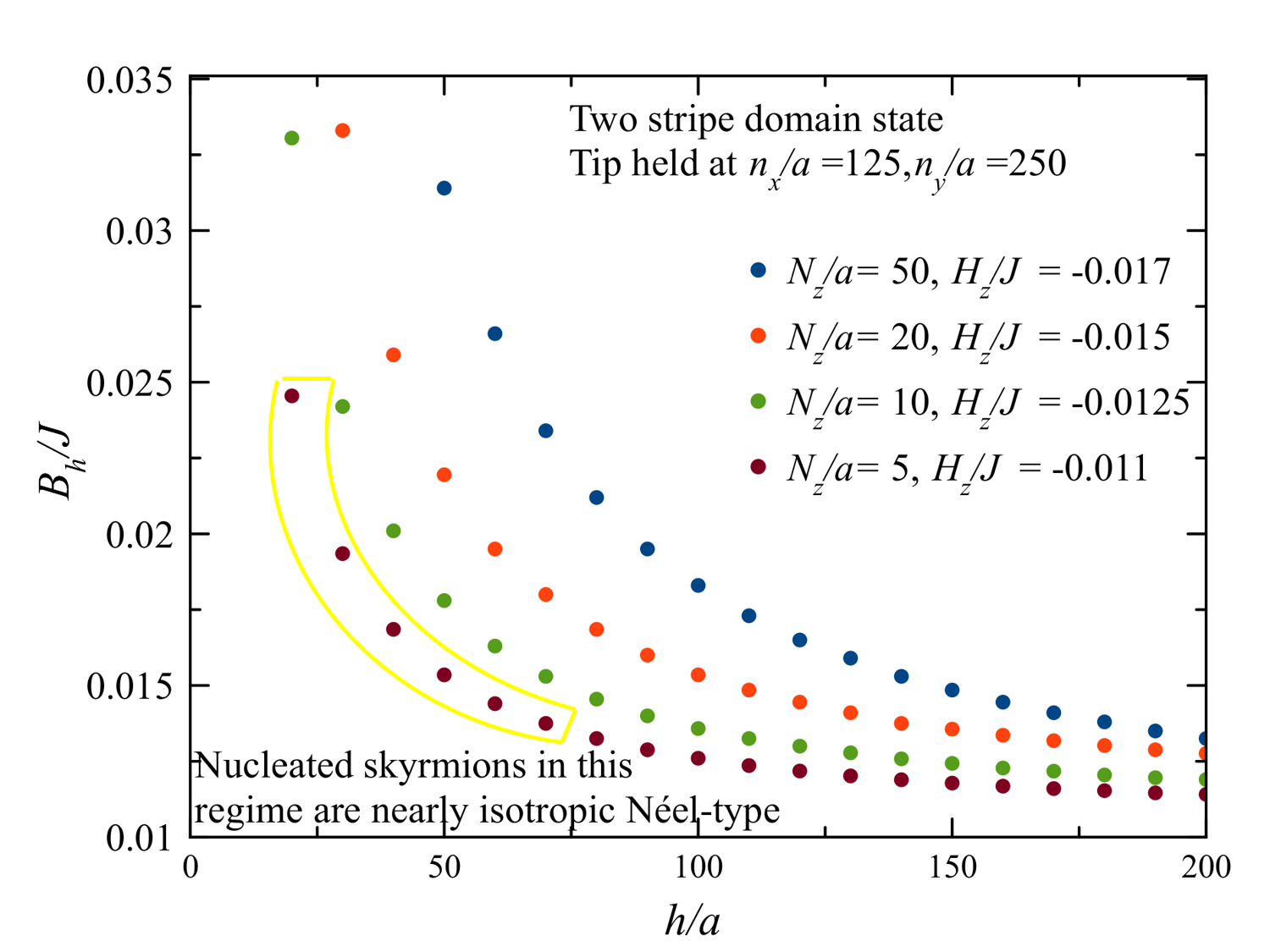}
\caption{Minimum tip field strength $B_{h}/J$ needed to nucleate the elliptical skyrmion as a function of tip height. The highlighted regime is where nearly isotropic N\'eel-type skyrmions were nucleated. These cannot be stabilized in the absence of the magnetic tip. }
\label{TipGraph}
\end{figure}

Next we attempted to see the affect of changing the film thickness on the nucleation process, again using the same initial state of the two vertical stripe domains. The process to obtain the minimum tip strength needed to nucleate the elliptical skyrmion as a function of the tip height was identical to the one used to obtain the data for Fig. \ref{TipHz}. We found that $m =0.132 h^{2.56}$ when $N_{z}=50$, while $m =0.060 h^{2.70}$, $m= 0.031 h^{2.82}$ and $m=0.022 h^{2.87}$ when $N_{z}=20$, $N_{z}=10$ and $N_{z}=5$, respectively, see Fig. \ref{TipGraph}. From this, we conclude that a stronger magnetic moment is needed to nucleate the skyrmion in the system with a higher number of layers, which seems to be due to the fact that the nucleation is contingent upon the applied field being strong enough to induce the ferromagnetic state in the absence of the tip. Since the DDI is amplified by the number of atomic layers, weaker applied fields are needed for this to occur and consequently $m$ can be smaller. To the contrary, if each system was subjected to the same external field that could induce the ferromagnetic state for all $N_{z}$, the required magnetic moment for elliptical skyrmion nucleation for the system with the largest number of layers would be smallest, as the prefactors increase with the number of layers since they are related to the strength of the applied field. 

\begin{figure}[ht]
\hspace{-0.5cm}
\centering
\includegraphics[width=6cm]{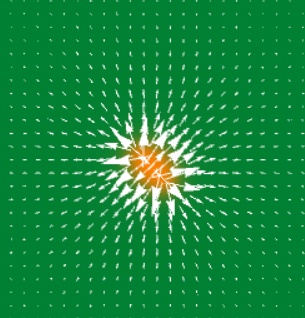}
\caption{N\'eel-type skyrmion nucleated by a magnetic tip when $h/a$=50, $H_{z}/J= - 0.011$, $B_{h}/J = 0.01537$. It is slightly deformed, hence the designation "nearly" isotropic. The numerically computed $\lambda_{x}$ is very close to $\lambda_{y}$. This is a cropped portion of the lattice. }
\label{Neel}
\end{figure}

This finding highlights the importance of the DDI in the elliptical skyrmion nucleation and stabilization. The effect of the DDI is amplified when there are more atomic layers, which requires a tip with a larger dipole moment, as well as a  stronger applied magnetic field. However, if the DDI is too weak, then the elliptical skyrmion will not be stable.  In particular, for the case $N_{z}=5$ for $h/a\leq 70$, the nucleated skyrmions were nearly isotropic N{\'e}el-type skyrmions rather than elliptical Bloch skyrmions, see Fig. \ref{Neel}. This seems to mark a transition region where it is increasingly difficult for the MFM to nucleate skyrmions altogether. We suspect the skyrmions are nucleated via brute force from the ferromagnetic state due to the tip field, like in \cite{GaraninMFM}, which is why they tend to have a chirality favored by the tip field.  Furthermore, we were unable to nucleate the skyrmions from the labyrinth domain state when $N_{z}=5$ because this is achieved when the tip is held closer to the film within the aforementioned height regime.  

\begin{figure}[ht]
\hspace{-0.5cm}
\centering
\includegraphics[width=8cm]{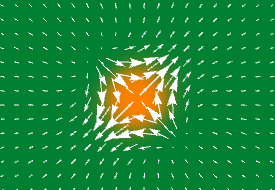}
\caption{Isotropic antiskyrmion of chirality $\gamma=\pi/2$ nucleated by a MFM tip held at height $h/a=100$ for $H_{z}/J= - 0.007$, $B_{h}/J = 0.008$. This is a zoomed-in section of the lattice.}
\label{AntiSkyrmTip}
\end{figure}

In fact, for the system consisting of a single atomic layer $N_{z}=1$, elliptical skyrmions will not be nucleated. Instead, one will nucleate isotropic antiskyrmions. An isotropic antiskyrmion of chirality $\gamma=\pi/2$ can be created from the stripe domain state by the external field alone if it strong enough. However, it will have a size of only a few lattice spacings. If a larger antiskyrmion is desired, one can be created by the combination of an applied field and a magnetic tip. One such example is shown in Fig. \ref{AntiSkyrmTip}.

\begin{figure}[ht]
\hspace{-0.5cm}
\centering
\includegraphics[width=8cm]{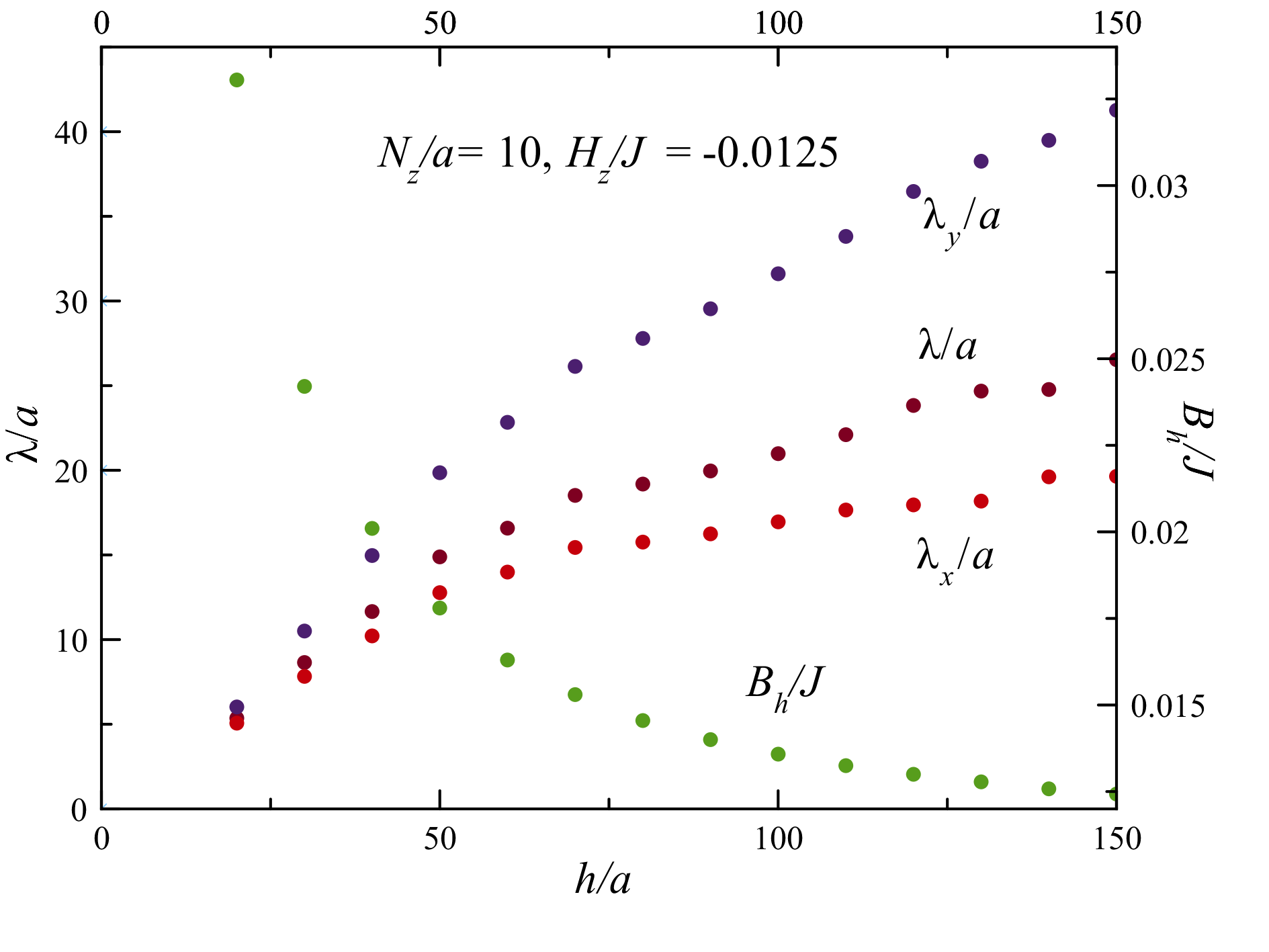}
\caption{Size of the nucleated elliptical skyrmion and the corresponding tip strength as a function of the tip height. $B_{h}(h)$ is the same as the curve in Fig. \ref{TipGraph} for $N_{z}=10$. The size $\lambda=\sqrt{\lambda_{x}\lambda_{y}}$ as well as the width and length $\lambda_{x}$ and $\lambda_{y}$ are computed using Eqs. (\ref{LamEff}) and Eqs. (\ref{LamEffx})}.
\label{lambda}
\end{figure}

Lastly, the size of the nucleated skyrmion also depends on the tip height. Namely, smaller skyrmions are nucleated at smaller tip heights. In a system where the tip field is the dominant interaction, one expects that the isotropic nucleated skyrmion should be of size $\lambda \sim h$, see Appendix \ref{Tip Analytics}. Instead, in the discrete lattice model with all relevant interactions, the behavior of  the skyrmion size is more complicated, as it depends upon $B_{h}$ and the applied field, in addition to the other material parameters, see Fig. \ref{lambda}. The decrease in the skyrmion size with decreasing $B_{h}$ involves a narrowing of the skyrmion as well as a decrease in length, i.e. both $\lambda_{x}$ and $\lambda_{y}$ decrease simultaneously. 

Ultimately, the skyrmion nucleation depends upon all controllable parameters in a potential experiment: the applied magnetic field, the strength of the magnetic tip, and the thickness of the film. In the material with the $D_{2d}$ crystal symmetry, the DMI favors elliptical skyrmions, see Eq. (\ref{EllipticalDMIEnergy}). The tip field also favors elliptical skyrmions, provided that the nucleated skyrmion is not $\lambda \sim h$, see Appendix \ref{Tip Analytics}, which was satisified for all cases we studied. Lastly, the nucleation depends on the film thickness, as the DDI scales with the number of atomic layers $N_{z}$. The film thickness can be used as a control parameter to determine the applied magnetic field and magnetic tip one wants to use in a real experiment. The only limitation is that if the film is too thin, the elliptical skyrmion will not be nucleated.

\section{Conclusions} \label{Sec_Conclusion}
We have shown that isolated elliptical skyrmions can be nucleated from a magnetic force microscope tip in a material with the $D_{2d}$ crystal symmetry which typically favors isotropic antiskyrmions using the real material parameters of Mn$_{1.4}$Pt$_{0.9}$Pd$_{0.1}$Sn. We find that the elliptical skyrmion can be nucleated from the stripe domain state or labyrinth domain state if one applies a magnetic field that is strong enough to destroy the domain state in conjunction with a magnetic tip held relatively close to the film, directly under the relevant section of stripe domain, with a tip field strength of the order of the applied field.

The nucleation of the elliptical skyrmion depends on the applied magnetic field in the sense that it must be strong enough to induce the ferromagnetic state. Additionally, the nucleation process depends  on the film thickness, which affects the strength of the magnetic tip required to nucleate the skyrmion. For the lattice spacing used in this paper, this seems possible in a film as thin as 25-50 nm. However, if the film is too thin, no elliptical skyrmions will be nucleated. As the strength of the DDI scales with the number of layers, this suggests that the elliptical skyrmion's nucleation and stability depend on a complicated interplay between the DDI and the DMI, which tends to favor antiskyrmions. In fact, we show that for a single atomic layer, the magnetic tip can be used to aid in the nucleation of antiskyrmions instead. 

We predict that materials with a DMI that supports isotropic skyrmions should also support elliptical antiskyrmions. A natural starting point to discover elliptical skyrmions experimentally would be to study materials where the DMI strength is of the order of the PMA strength and the ratio of the PMA to the DDI is of the order unity.  Such a system would support the same domain states studied in this paper. Presumably elliptical antiskyrmions could be nucleated using a MFM in this case. 

Counterclockwise or clockwise Bloch skyrmions can be nucleated from vertical or horizontal stripe domains, respectively.  The chirality (rotation of the in-plane spin components), as well as the length and width of the elliptical skyrmion can be used as additional degrees of freedom for potential computing applications. It is also noteworthy that a chiral material can support both states, as the DMI typically supports a single rotation of the spins.

Finally, the magnetic moment of the elliptical skyrmion is similar to the isotropic skyrmion and also depends on both of these characteristic lengths. Elliptical skyrmions, which seem to move faster than isotropic skyrmions in response to an applied current, are characterized by two length scales, and have two potential rotations of the in-plane spins open new and exciting possibilities for potential computing applications.

\section{Acknowledgements}

The author thanks E. M. Chudnovsky for reading the manuscript and D. A. Garanin for helpful discussions.

 \appendix

\section{Tip Analytics} \label{Tip Analytics}
One can compute the Zeeman energy of the BP skyrmion due to the magnetic tip in the continuous limit for the most general case of the elliptical skyrmion spin field using
\begin{equation}
E_{tip}= -\frac{1}{a^2}\int d^2\rho\;\left( \mathbf{B}_{tip} \cdot \mathbf{s} + B_{tip,z}\right).
\end{equation}
In the above, we subtract the contribution due to the uniform background where $\mathbf{s} = -\hat{\mathbf{e}}_{z}$. 

Upon substituting Eqs. (\ref{SEllip}) and Eq. (\ref{TipField}) and integrating over the angle, one obtains:
\begin{equation}
E_{tip}= -\frac{2\pi B_{h} h^3 \lambda_{x}\lambda_{y}}{a^2} I,
\end{equation}
where
\begin{eqnarray}
I&=&  \int_{0}^{\infty} \rho d\rho\Bigg.\Bigg\{ \frac{(2h^2-\rho^2)}{(\rho^2+h^2)^{5/2}\sqrt{(\rho^2+\lambda_{x}^2)(\rho^2+\lambda_{y}^2)}} \\&+&\frac{3h\left(\rho^2-\lambda_{x}\lambda_{y}+\sqrt{(\rho^2+\lambda_{x}^2)(\rho^2+\lambda_{y}^2)}\right)\cos\gamma}{(\lambda_{x}+\lambda_{y})(\rho^2+h^2)^{5/2}\sqrt{(\rho^2+\lambda_{x}^2)(\rho^2+\lambda_{y}^2)}}\Bigg.\Bigg\}. \nonumber \label{TipIntegral}
\end{eqnarray}
For fixed tip strength $B_{h}$, the global minimum occurs at $\lambda_{x}=\lambda_{y} = 1.41 h$ for the chirality $\gamma=0$ N\'eel-type skyrmion, as calculated in \cite{CapicInteraction}. 

\begin{figure}[ht]
\hspace{-0.5cm}
\centering
\includegraphics[width=8cm]{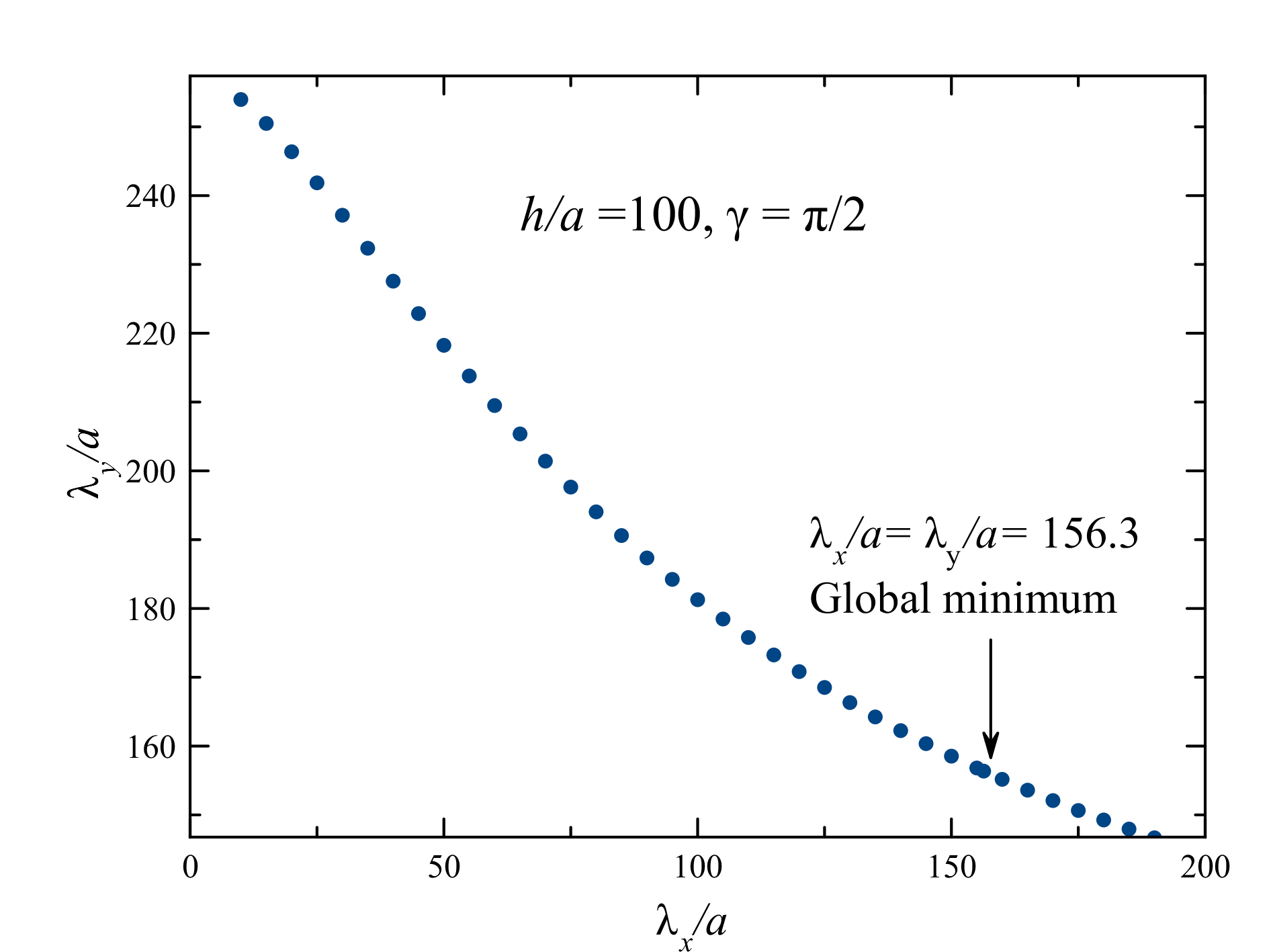}
\caption{$(\lambda_{x}, \lambda_{y})$ pairs that minimize the energy in Eq. (\ref{TipIntegral}) for fixed  $h$.}
\label{lamTip}
\end{figure}

Relevant to this paper, if one does the analysis of $I$ at fixed skyrmion width $\lambda_{x}$ for example, the energy is minimized when $\lambda_{y}\neq \lambda_{x}$. This is true for all cases except for the global energy minimum which occurs when $\lambda_{x}/a =\lambda_{y}/a=1.563 h/a$ for the Bloch-type skyrmion,  see Fig. \ref{lamTip}. Therefore, in the presence of other interactions that favor the elongation of the skyrmion, the metastable elliptical skyrmions also have $\lambda_{y}\neq \lambda_{x}$.

\pagebreak

\end{document}